\documentclass[ times,1p,twocolumn,sort&compress ]{elsarticle}

\usepackage{graphicx}
\usepackage{caption}
\usepackage{subcaption}
\usepackage{amssymb,amsmath}
\usepackage[mathscr]{eucal}

\begin{document}

\begin{frontmatter} 
\title{Spiral orbits and oscillations in historical evolution of empires}

\author{Taksu Cheon}
\ead[url]{http://researchmap.jp/T\_Zen/} 
\author{Sergey S. Poghosyan}
\ead[url]{http://researchmap.jp/sergey.poghosyan/} 

\address{
Laboratory of Physics, Kochi University of Technology, Tosa Yamada, 
Kochi 782-8502, Japan
\\
\bigskip
{\rm To the memory of Kijae Cheon (1930--2016)}
} 
%
\date{\today}
%
%
\begin{abstract}
We introduce the concept of metaasabiya, the second non-material resource, to the asabiya theory of historical dynamics.  We find that the resulting three variable dynamical system has peculiar features such as repelling or attracting axes and spiralling orbits in the phase space. Depending on the initial state, the system can go through series of oscillatory rises and falls, mimicking the geopolitical evolution of real-world polities.  These distinctive features, absent in conventional Lotka-Volterra type biological systems,  reveal the hidden richness inherent in the asabiya theory.
\\
\end{abstract}


\end{frontmatter} 

\section{Introduction}

The success of mathematical biology has been so impressive that we naturally expect the emergence of the mathematical theory of the evolution of human societies.  The expectation has been met with the asabiya theory of Turchin \cite{TU03, TU08}, who has written down a set of dynamical equations that describe the temporal evolution of  the size of a polity and {\it asabiya}, its non-material resource, a concept adopted from the classic study of medieval North African history by Ibn Khaldun, representing the social cohesion that plays the key role in the growth and the decay of a polity.

The set of equations turns out to be capable of describing a rise and fall of a polity thanks to the presence of a fixed point of the system which acts as a repeller in the phase space.   
Turchin then considered several polities, each described by its size and asabiya, interacting through  a geographically connected network of polities, in similar fashion to earlier attempts based on the ``geopolitical'' theories \cite{AK96}.  
The evolution of polities has shown intricate patterns of oscillations that represent repeated rise and fall of polities, which could even be made to resemble the real historical data, with proper tuning of parameters \cite{TH03}. 
We can, then, pose a question whether the historical cycle of a nation's rise and fall is necessarily the result of their mutual interactions.   Is it that a polity can rise and fall only once, if it is left more or less undisturbed by other competing polity?  That is unlikely.  It rather seems reasonable to us, that there is an internal mechanism for the oscillating fortunes of polities, even in the absence of its interaction with others, and there should be a set of equations that, by itself alone, could capture the essence of dynamical oscillations found in human societies \cite{TN08}.

In this article, we present one candidate for such models with an introduction of the second non-material resource, which we call {\it metaasabiya} to the Turchin's equation.   
The concept of metaasabiya is a mathematical implementation of the historical philosophy described in the works of Oswald Spengler \cite{SP18} and Arnold Toynbee \cite{TO34}, who had both stressed the cultural dimension of the sustenance of a polity in its later stage.  They had observed that the polities, after the waning of their initial impulse for the creative evolution, can still resort to an institutionalized ideological system -- morphism of {\it Kultur} into {\it Zivilisation} in Spengler's terminology, and the emergence of the {\it universal state} and the {\it universal religion} in Toynbee's terminology -- for their second stage growth. Indeed, it is generally acknowledged that institutionalized religion with its derivatives in the form of scientific learning always accompany the great rise of nations in its later ``imperial'' stage.  If we are to quantify the effect of the religion, it should naturally be considered as a quantity separate from asabiya, a quantity which can rise after the emergence of asabiya.  This is exactly the reason for our choice of the neologism ``metaasabiya'' to represent this quantity.

Interestingly, the reexamination of Ibn Khaldun's text reveals that the embryonic form of the concept of metaasabiya, as we have outlined, can already be found in his writings.  He had described the the quick emergence of elaborate court culture in the Arab empire after its great conquest of Mesopotamia, with detailed analysis of its effects, both positive as well as negative, on the social cohesion of the conquerer's polity. 

In the following sections, we detail our population dynamical model with asabiya and metaasabiya (section 2), pin down the skeleton of its dynamics in terms of the invariant surface, and analyze the orbits in the  phase space to find spiral motion around the repeller (section 3), study the nature of the spiral by linearised map around the repeller (section 4), and look into the structure of basin of attraction (section 5). 
We end the article with a discussion (section 6) on the implication of our findings and possible future direction of the study of our model.    

\section{Population dynamics with asabiya and metaasabiya}

We consider a  {\it polity}, a socio-political entity made up of people, such as a tribe, a city-state, a nation, which we characterize by three time-dependent quantities the size $A(t)$, the asabiya $S(t)$, and the metaasabiya $R(t)$. 
We can identify the size of a polity either with its population or by the territorial area it controls, or alternatively, by its economic strength measured, for example, in terms of its gross domestic product (GDP).  
The asabiya and the metaasabiya are both non-material ``spiritual'' resources of a polity that bind people together and help the society to grow.  The asabiya, as considered by Ibn Khaldun and by Turchin, represents the social cohesion of a polity that helps to increase its growth through higher birth rate, through the absorption of surrounding ``barbarian'' population, and through the military victories over rival polities.  When the polity's asabiya is lost, often  as a result of overabundance of elite population and loss of fiscal discipline, the growth rate of polity decreases and even become negative.  Metaasabiya represents ''cultural'' assets of the polity, such as religion, science, art and entertainment, that work positively for the growth of a polity both by easing its internal tensions and providing technological advances.  Metaasabiya, understood in this way, can also directly contribute to the increase of the polity's size by assimilating foreign elements in and around peacefully.  

We consider the set of evolution equations given by
\begin{eqnarray}
\label{exturchin}
&&
\frac{d A}{d t} = \frac{c}{1+f}(S+f R) A \left( 1-\frac{1}{h} A\right) - a \, ,
\nonumber \\
&&
\frac{d S}{d t} = r \left( 1- \frac{1}{2b} A\right) S (1-S) \, ,
\nonumber \\
&&
\frac{d R}{d t} = q \left( 1- \frac{1}{2d} A\right) S R (1-R) \, .
\end{eqnarray}
In the first equation, the term $A(1-A/h)$ describes the logistic evolution of the polity size $A$ whose growth limit is given by the quantity $h$ determined by the natural resource of the land.  The factor $S+fR$ signifies the necessity of the presence of either the asabiya $S$ or the metaasabitya $R$ as the condition for the growth, where the parameter $f$ gives the ratio of respective contributions.  The second term $-a$ gives the threshold, and when the first term is below $a$, population declines.  

The parameters $b$ and $d$, in the second and third equations respectively, are the key quantities of the system:
The asabiya $S$, that quantify the internal cohesion of the polity takes the value $0 \leqslant S \leqslant 1$.  The growth rate of asabiya is higher for smaller size of the population and it becomes zero at the critical population $A=2b$.  The metaasabiya $R$, a new concept of this article, also takes the value $0 \leqslant R \leqslant 1$.  It quantifies the cultural resource of the polity.  The law governing the growth of metaasabiya is similar to that of asabiya (with the critical population 2d) except that the presence of larger value of asabiya is the prerequisite for the growth of metaasabiya.  We can naturally assume that the saturation point of cultural assets, the metaasabiya $2d$ is higher than that for social cohesion, the asabiya $2b$.  In this article, we mainly consider the case of $2b$ $<\frac{h}{2}$ $< 2d$, which turns out to produce most interesting results. 
The parameters $c$, $r$ and $q$ gives the scale for the growth rates of $A$, $S$ and $R$ respectively.
 These equations can be thought of as representing Ibn-Khaldunian and Toynbeean view of history made from environmental challenge $-a$ met by human ingenuities $S+fR$ which decrease as a result of success.

 In the absence of the metaasabiya, namely, with $R=0$, we obtain the Turchin's cliodynamical equation that couple the population $A(t)$ and the asabiya $S(t)$;
\begin{eqnarray}
\label{orturchin}
&&
\frac{d A}{d t} = c S A \left( 1-\frac{1}{h} A\right) - a \, ,
\nonumber \\
&&
\frac{d S}{d t} = r \left( 1- \frac{1}{2b} A\right) S (1-S) \, .
\end{eqnarray}
It is shown that this set of equations, unlike typical population dynamical equations in mathematical biology, can possess unstable spiral fixed point, with proper tuning of parameters, and can describe a single ``rise and fall'' of $A(t)$ that characterizes the temporal evolution of historical human political organizations \cite{TU03}.

\begin{figure}[h]
\center
  \includegraphics[width=5.5cm]{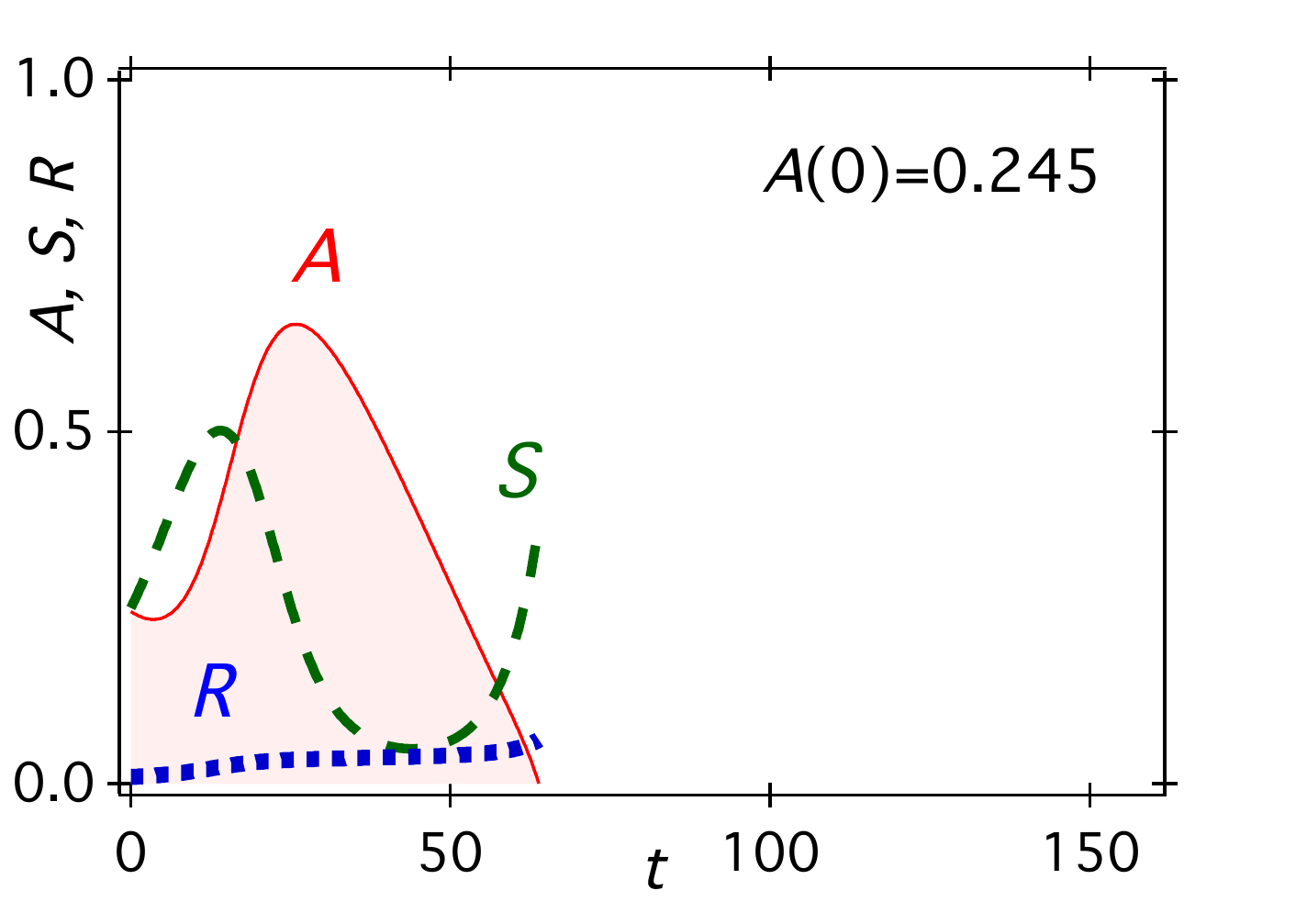}
  \includegraphics[width=5.5cm]{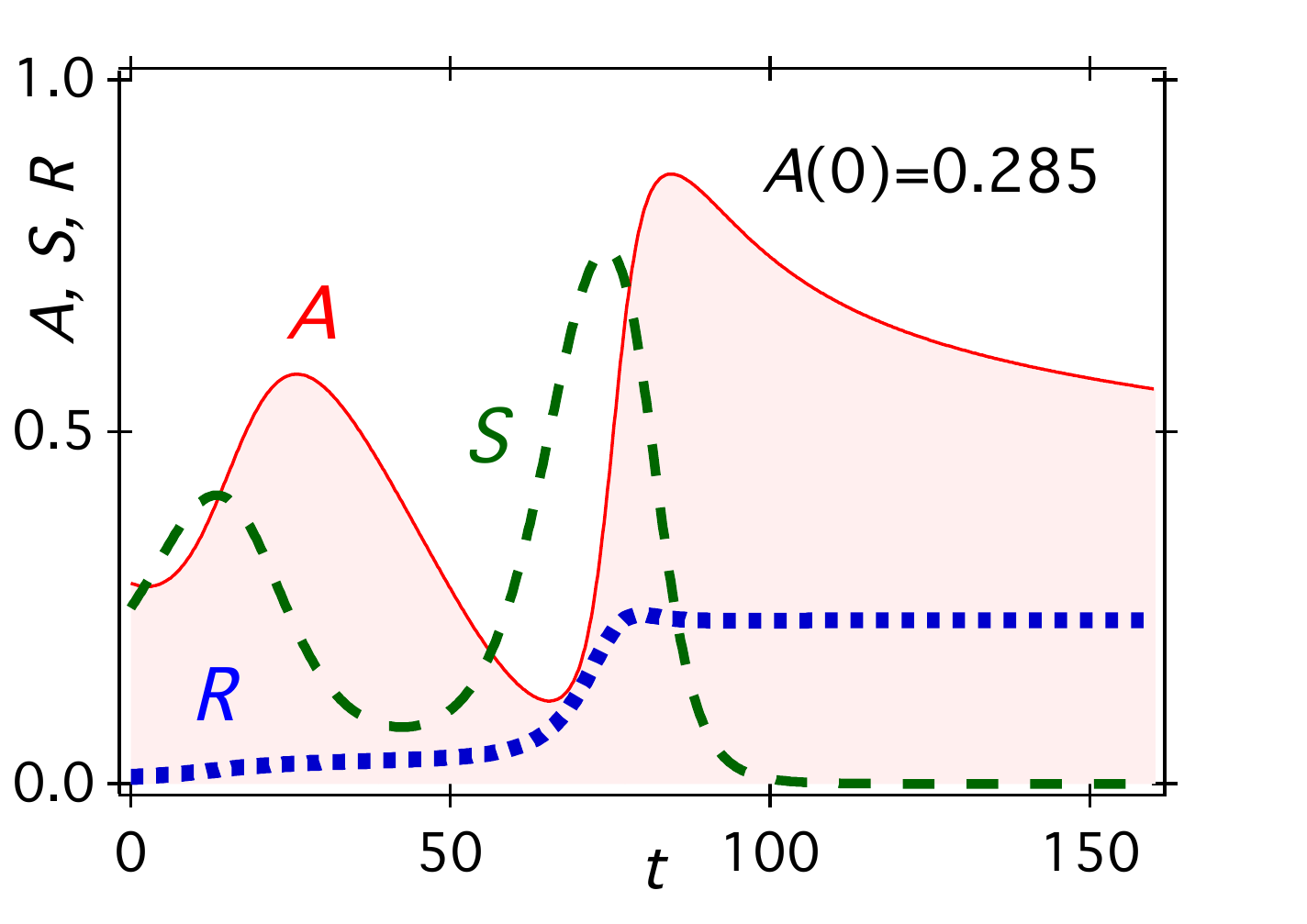}
  \\
\center
  \includegraphics[width=5.5cm]{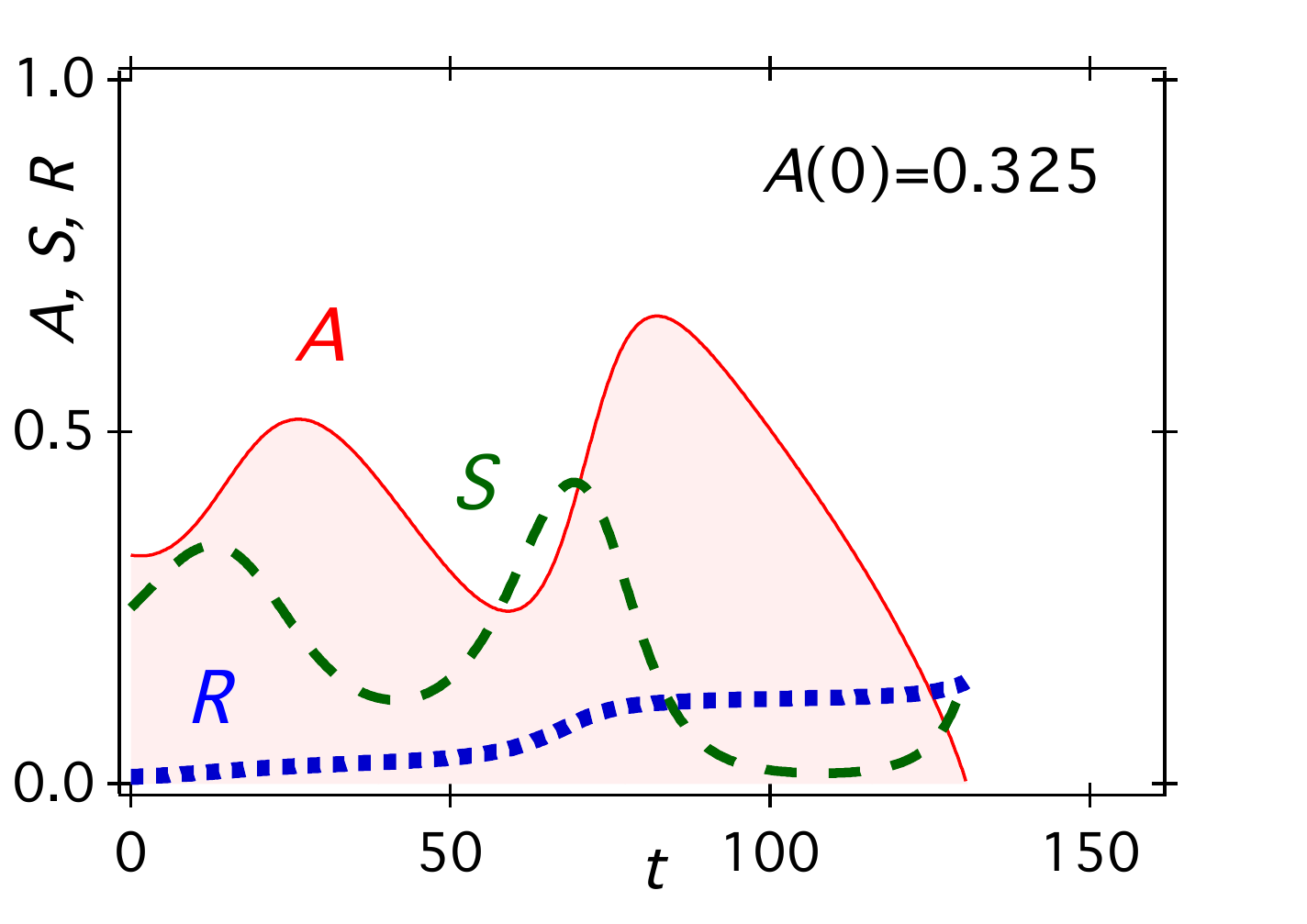}
  \includegraphics[width=5.5cm]{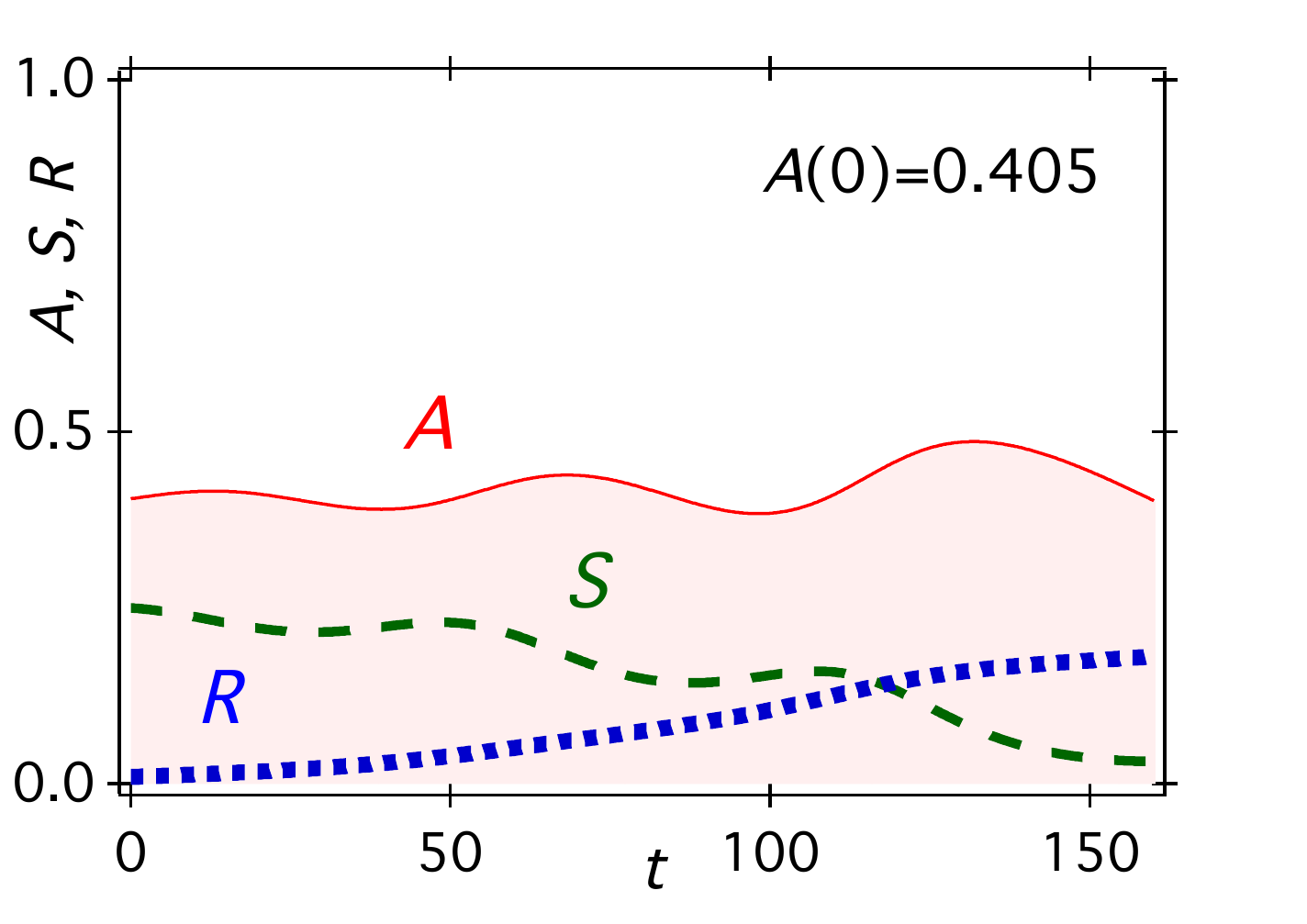}
\caption{The solutions of (\ref{exturchin}) representing the temporal evolution of the size $A$ (solid red lines), the asabiya $S$ (dashed green lines), and the metaasabiya $R$ (dotted blue lines), 
starting from the initials values $A(0) = 0.245$ (top left),
$A(0) = 0.285$ (top right), $A(0) = 0.325$ (bottom left), and $A(0) = 0.405$ (bottom right),
describing the historical evolution of a polity with a single rise and fall, a permanence after two oscillations, a fall after two oscillations, and a stable permanence, respectively.  Other initial values are set to be $S(0)=0.25$ and $R(0) = 0.05$.
The system parameters are set to be $c=1$, $f=1$, $h=1$, $a=\frac{3}{100}$, $r=q=\frac{1}{4}$, $b=\frac{1}{5}$ and $d = \frac{2}{5}$.
}  
\label{f1}
\end{figure}
We now show that our set of equation (\ref{exturchin}), with the introduction of the third dynamical variable $R$, can display far richer dynamics that mimics the temporal evolution of real-world polities.  
Let us start by showing some representative numerical solutions of the equation (\ref{exturchin}) in Fig. 1. 
We observe, in Fig. 1, that depending on the initial values of $A(0)$, $S(0)$, and $R(0)$, the polity can experience a single rise and fall (a), an initial rise and fall followed by the second rise and fall (b), rise and fall and a second rise followed by a convergence to a stable state (c).  It can also stabilize after eventless small oscillation (d).

An early rise of the population is always preceded by a high value of asabiya, and likewise, the falling asabiya is a precursor to  the fall of population.  The metaasabiya usually increases slowly, but experiences a sudden increase in time of trouble in which falling population is combined with surging asabiya.
While initial higher value of asabiya $S$ propels the polity to an initial rise, the rise itself brings about the decrease of asabiya which eventually leads to the decline of the polity's population.  With the presence of metaasabiya $R$ that starts to increase following the increase of asabiya $S$, it is sometimes possible that the population starts to grow for the second time, and the polity can experience the ``imperial rise'' that is mostly supported by the metaasabiya $R$ rather than by the asabiya $S$.   
With proper sets of parameters and initial conditions, it is even possible that the population reaches a stable nonzero value, after several oscillations, signifying the polity's ``millennium empire'' status, which represents such polities as Ancient Egypt and Byzantine empire which had lasted for the time scale far beyond the normal life span of few hundred years.
These features all  agree very well with our knowledge on the evolution of historical polities, kingdoms and empires. 
%

\section{Invariant surfaces, attractor and repeller}

\begin{figure}[h]
\centering
\includegraphics[width=4.8cm]{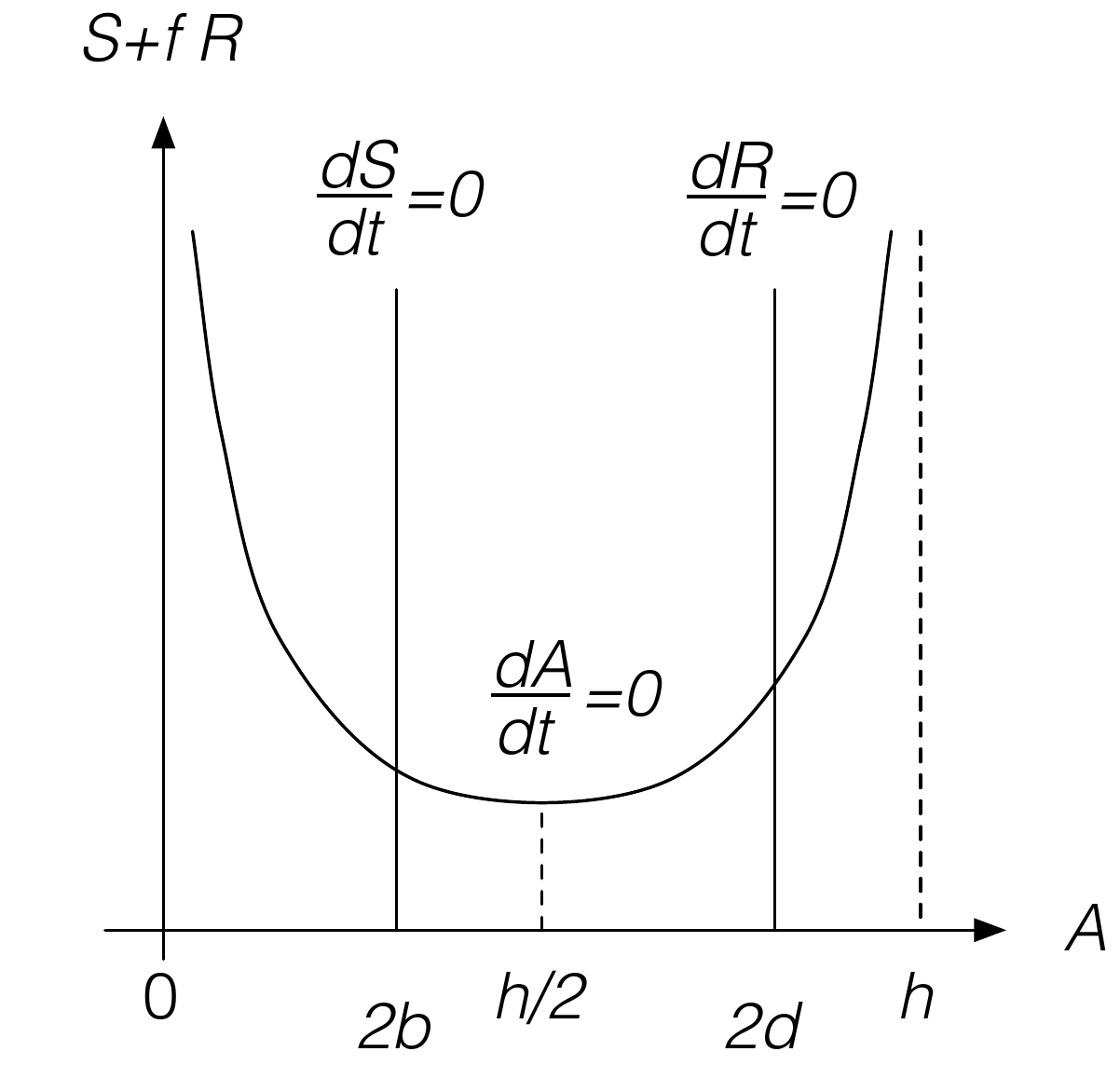}
%
\caption{Isoclines of the dynamical system (\ref{exturchin}), speified by $\frac{d A}{d t}=0$, $\frac{d S}{d t}=0$, and $\frac{d R}{d t}=0$.}
\label{f2}
\end{figure}

In this and the next sections, we explore the behind-the-scene reason of the dynamical richness of systems described by the equation (\ref{exturchin}).
The phase space structure of the  system of coupled nonlinear differential equation (\ref{exturchin}) is readily analyzed with standard tools of dynamical systems theory \cite{AS96}.
The skeleton of the phase space structures are given by the fixed points of the system, which are obtained as the conjunction of isoclies 
$\frac{d A}{dt}=0$, $\frac{d S}{dt}=0$, and $\frac{d R}{dt}=0$ which are depicted in Fig. \ref{f2}.  We treat the special case of $b=d$ first, and consider the case of our prime interest, $2b<h/2 < 2d$ later on.
%

\subsection{The case  $b=d$}

Three sets of fixed point lying on the surface of unit cube are easily found.
\begin{itemize}
\item 
Four points specified by 
\begin{eqnarray}
&& 
S=1, \quad R=0, \quad A=\frac{h}{2} \left( 1 \pm \sqrt{1-\frac{8a}{hc}} \right), 
\nonumber \\   
&& 
{\rm and \ by}
\nonumber \\
&&
S=1, \quad R=1, \quad A=\frac{h}{2} \left( 1 \pm \sqrt{1-\frac{4a}{hc}} \right) .
\label{fxtrv}
\end{eqnarray}
These fixed points are hyperbolic.
\item The curve specified by 
\begin{eqnarray}
S=0, \quad R = \frac{a(1+f)}{c f A (1-A/h)} .
\label{fxr0}
\end{eqnarray}
Closer examination with the linearization of the evolution equation around the points on the curve reveals that all points on the curve satisfying $A>2b$ are attractors of the system.  Likewise, all points on the curve satisfying $A<2b$ are repellers.
\item A point specified by
\begin{eqnarray}
R=0, \quad S=S^\star, \quad A=2b ,
\end{eqnarray}
where $S^\star$ is defined by
\begin{eqnarray}
S^\star = \frac{a(1+f)h}{2 c b (h-2b)} .
\label{fxs0}
\end{eqnarray}
This is an attractor if we have $2b > \frac{h}{2}$, and a repeller in the case $2b < \frac{h}{2}$.
\end{itemize}
\begin{figure}[h]
\centering
\includegraphics[width=4.2cm]{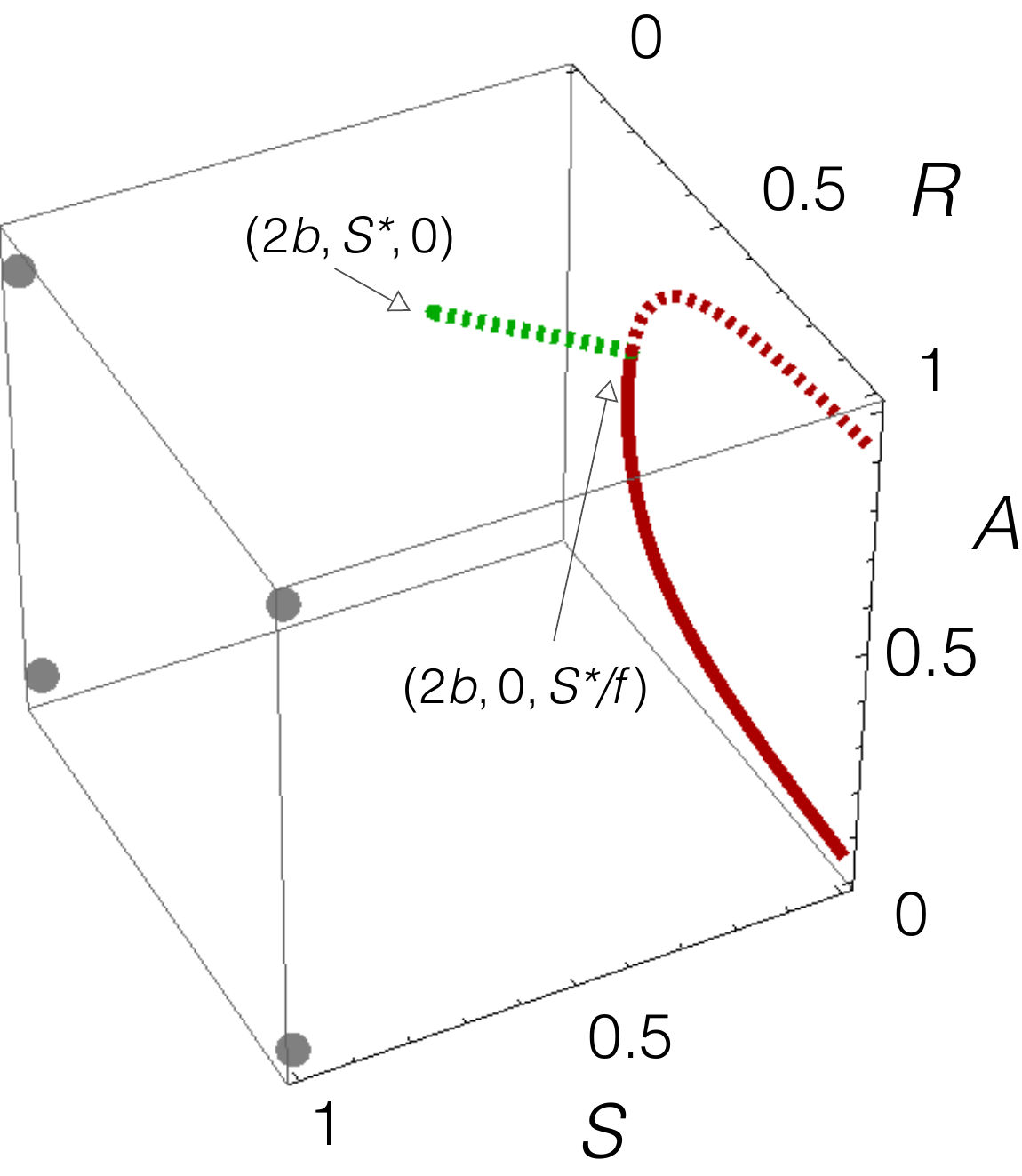}
%
\qquad\qquad
\includegraphics[width=4.2cm]{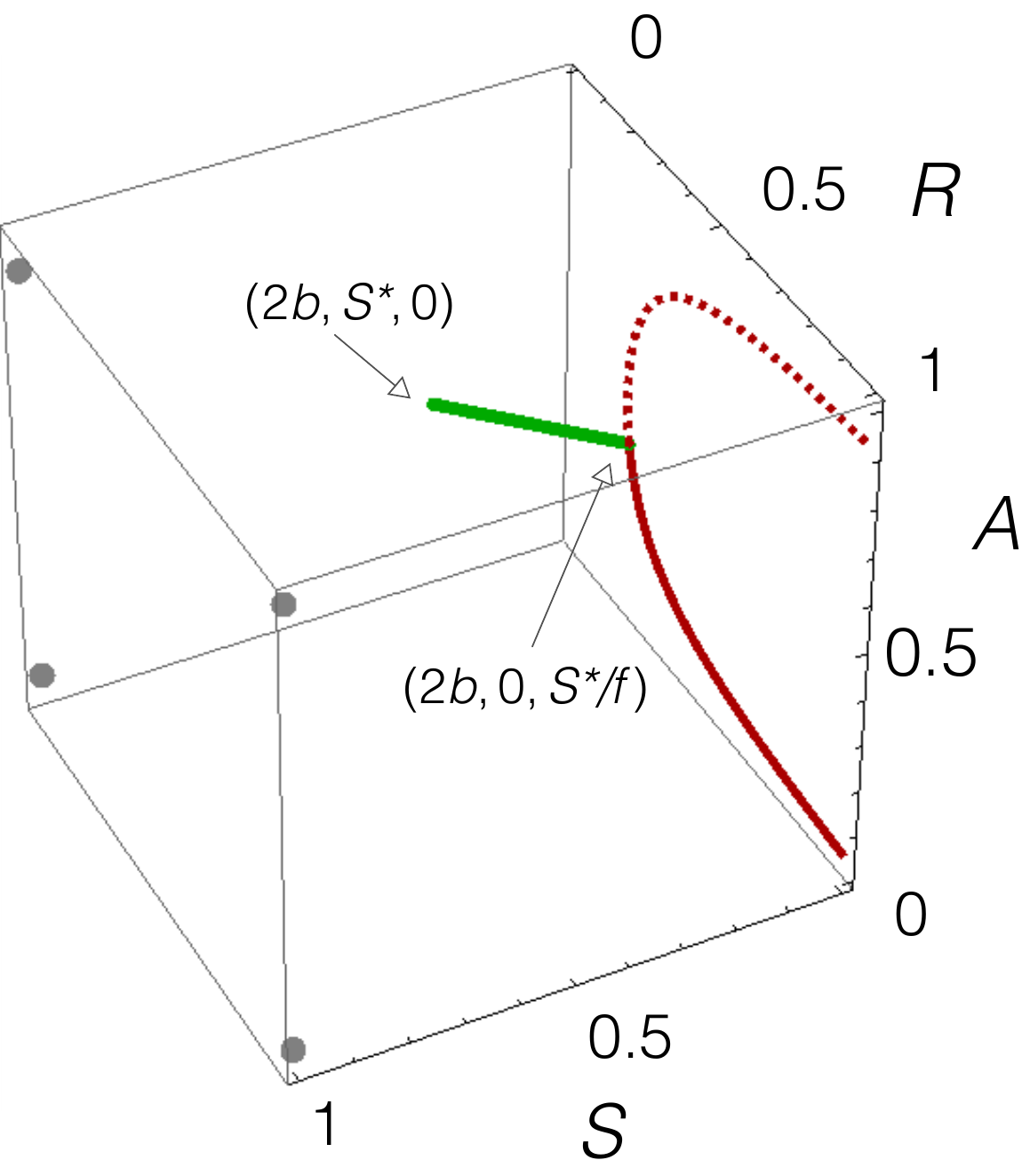}
\caption{Invariant surfaces of the system with $b=d$.  The grey points are the hyperbolic fixed points. The solid and dotted red lines represent dense collections of repulsive and attractive fixed points, respectively.  The green line is the invariant surface to which orbits are attracted (in case of dotted lines), or from which they are repulsed (in case of solid lines).  The figure on the left depicts the case of $2b=2d>\frac{h}{2}$ while the one on the right the case $2b=2d<\frac{h}{2}$.}
\label{f3}
\end{figure}
For the case of $b=d$, two isoclines for $\frac{dS}{dt}=0$ and $\frac{dR}{dt}=0$ merge, and we have nontrivial fixed points
given by their intersection with the isocline $\frac{dA}{dt}=0$ as
\begin{eqnarray}
S+fR=S^\star , \quad
A=2b .
\label{stableline}
\end{eqnarray}
%
This set of points represents a line whose one endpoint on $S=0$ plane coincides with the attractor-repeller dividing point on the curve (\ref{fxr0}), and another endpoint on $R=0$ plane coincides with the point (\ref{fxs0}). The points on (\ref{stableline}) are attractors if $2b>\frac{h}{2}$ and repellers if $2b<\frac{h}{2}$.   The situation is best  understood by inspecting Fig. 3, in which fixed points (\ref{fxtrv}) - (\ref{stableline}) are depicted on the unit cube representing $(S, R, A)$ with the choice $h=1$.

To analyze the trajectory of the system, we write
\begin{equation}
\frac{dS}{dR}=\frac{r}{q}\frac{1-S}{R(1-R)}.
\end{equation}
which is obtained from the second and the third lines of (\ref{exturchin}).
Setting initial conditions$A(0)=A_0$, $S(0)=S_0$, and $R(0)=R_0$, we derive an orbit of the evolution 
\begin{equation}
\frac{q}{r} \log{\frac{1-S}{1-S_0}}=\log{\frac{1-R}{1-R_0}} - \log{\frac{R}{R_0}} .
 \label{trajectory}
 \end{equation}
For the case of $2b > \frac{h}{2}$, trajectories either end up hitting $S=0$ surface, or absorbed to a point on the attractor (Fig. 4).
%
%
%
\begin{figure}[h]
\centering
\includegraphics[width=4.2cm]{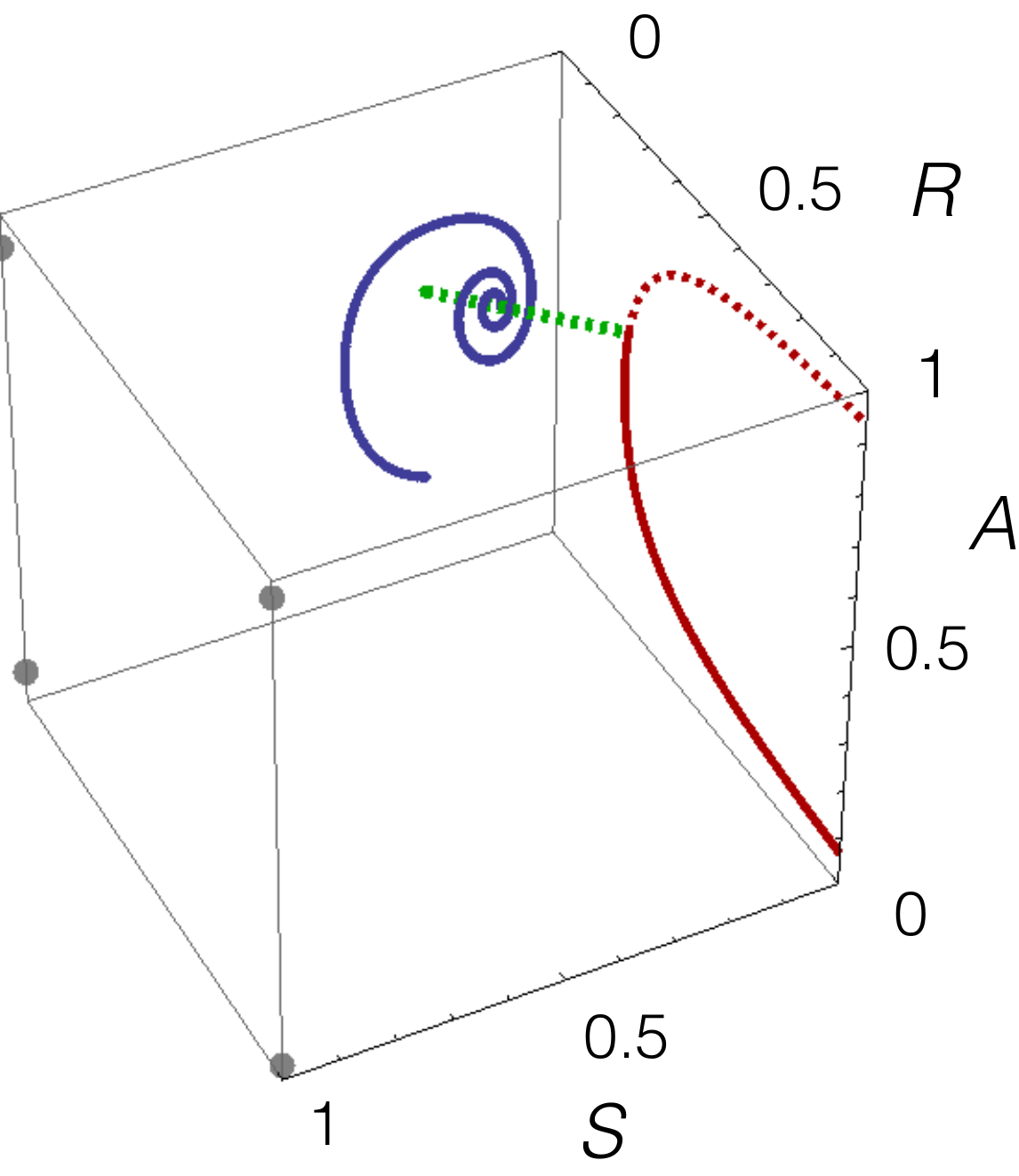}
%
\qquad\qquad
\includegraphics[width=4.2cm]{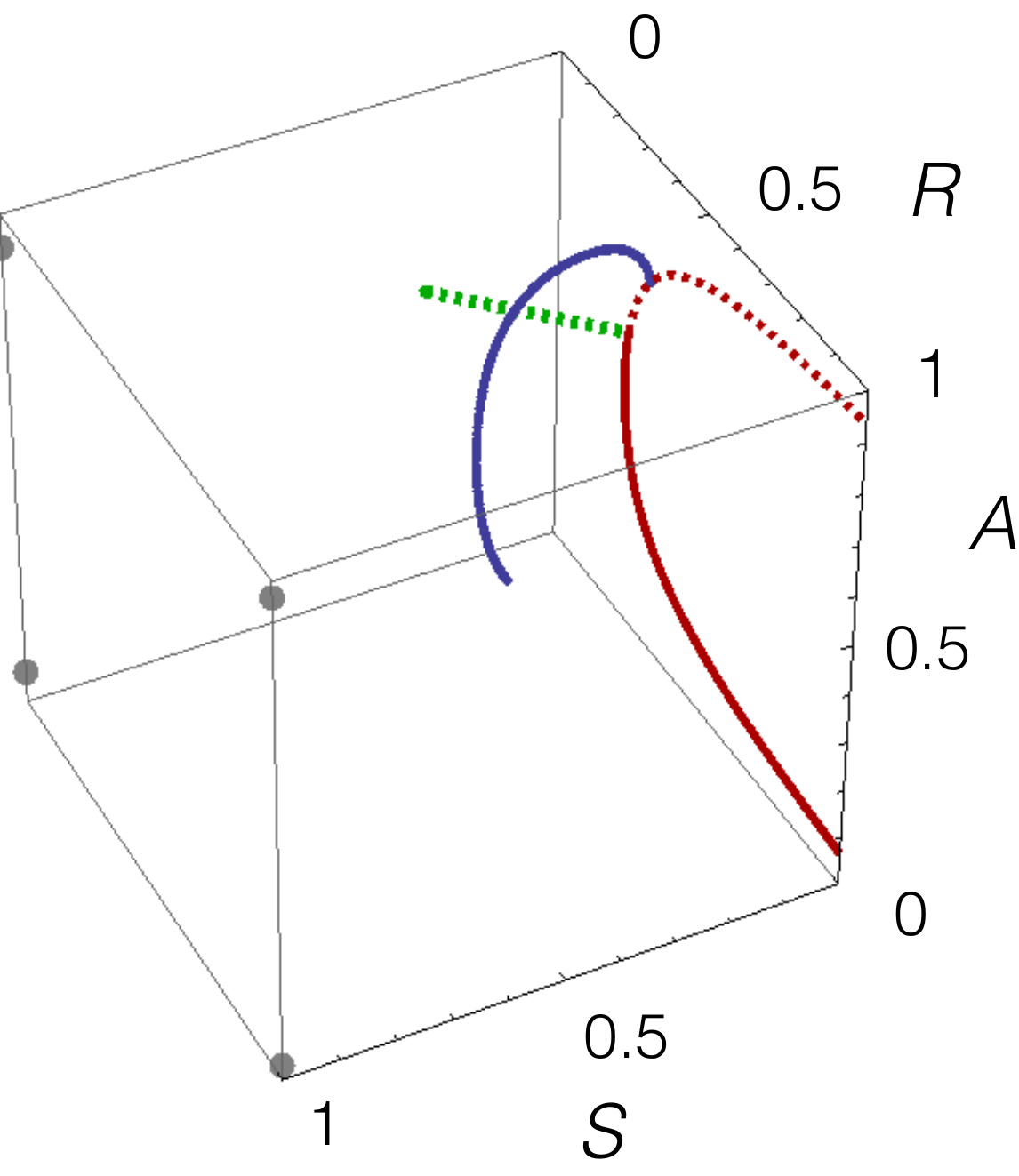}
\caption{The dynamics of the model for the case of the line (\ref{stableline}) being the attractor.  The blue line is a trajectory of polity evolution of the system with parameter values:
$h=1$, $c=1$, $f=1$, $b=d=0.3$, $r=q=0.25$, and  $a=0.03$.  The initial conditions are $S_0=0.3$, $R_0=0.1$ and $A_0=0.3$  for the figure on the left,  and  $S_0=0.3$, $R_0=0.4$ and $A_0=0.3$ for the right.}
\label{f4}
\end{figure}

Let us assume that the initial point of the trajectory is close enough to the attracting line, so the evolution of the system ends up on it (left case of Fig. 4). To estimate the final values of asabiya and metaasabiya, we assume that the initial points $(S_0, R_0)$ are close enough to equilibrium point $(S,R)$.  Substituting $S$ in (\ref{trajectory}) by $S=S^\star-fR$ obtained from (\ref{stableline}), and utilizing an approximation
$\log{(1+x)}\approx x$ for $|x|\ll1$, we have
%
\begin{eqnarray}
1-\frac{1-S^\star+fR}{1-S_0}=\frac{r}{q}\left(\frac{R}{R_0}-1-\frac{1-R}{1-R_0}+1\right).
\end{eqnarray}
We obtain an estimation for the final state of the system in the form
\begin{eqnarray}
&&
S(\infty)=S^\star-fR_0 \frac{ q(S^\star-S_0)(1-R_0) + r (1-S_0) } { fq R_0(1-R_0)+ r (1-S_0) },
\nonumber \\
&&
R(\infty)=R_0\frac{ q(S^\star-S_0)(1-R_0) + r (1-S_0) } { fq R_0(1-R_0)+ r (1-S_0) },
\nonumber \\
&&
A(\infty) = 2b,
\label{finalvaluesexact}
\end{eqnarray}
which agrees well with the numerical calculation.

For the case of $2b < \frac{h}{2}$, the trajectories starting with $R_0\lesssim S^\star/f$ spiral around the repeller and end up hitting $A=0$ surface, and the ones starting with $R_0\lesssim S^\star/f$ gets absorbed into the attractor (Fig. 5).
\begin{figure}[h]
\centering
\includegraphics[width=4.2cm]{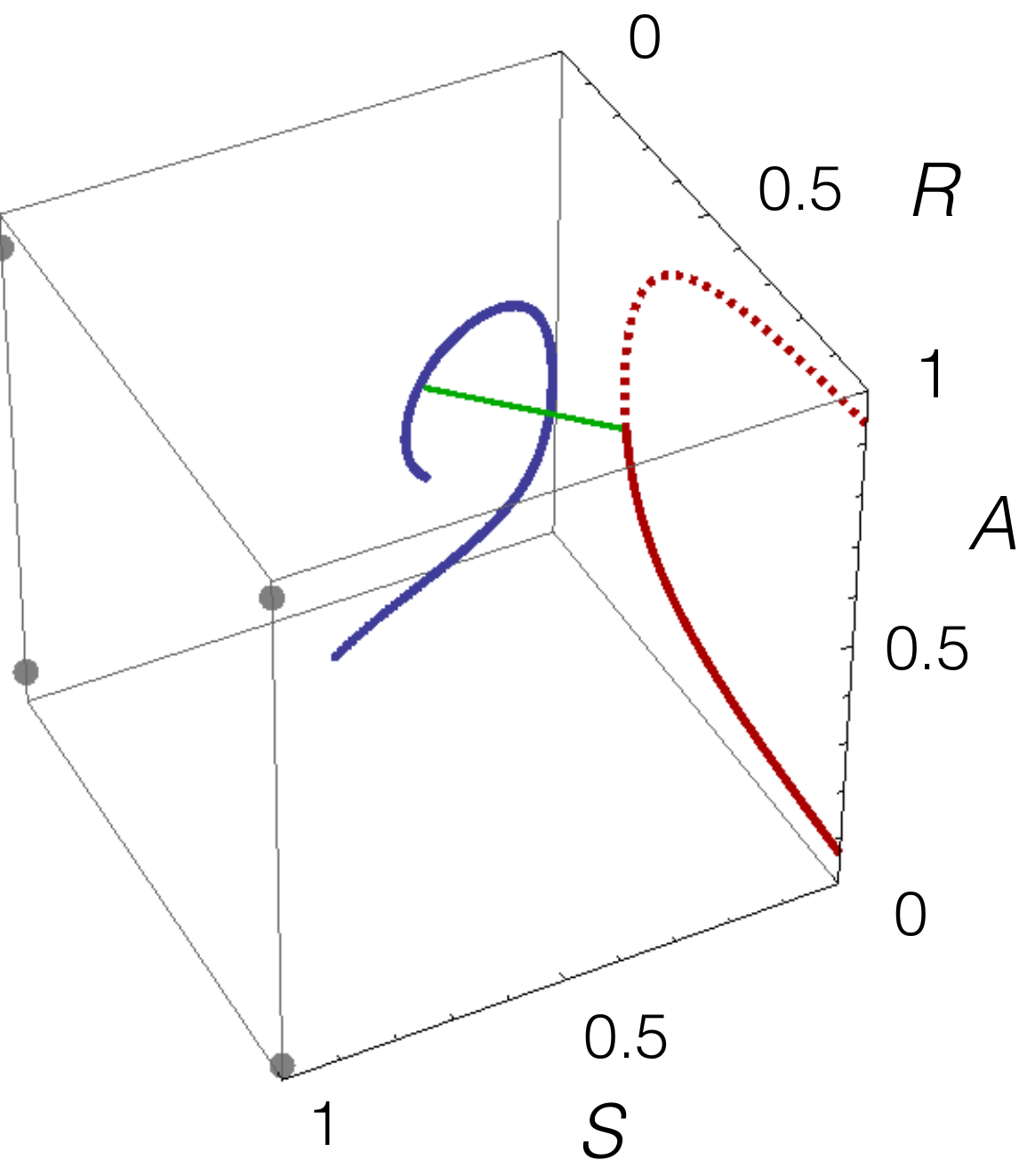}
\qquad\qquad
\includegraphics[width=4.2cm]{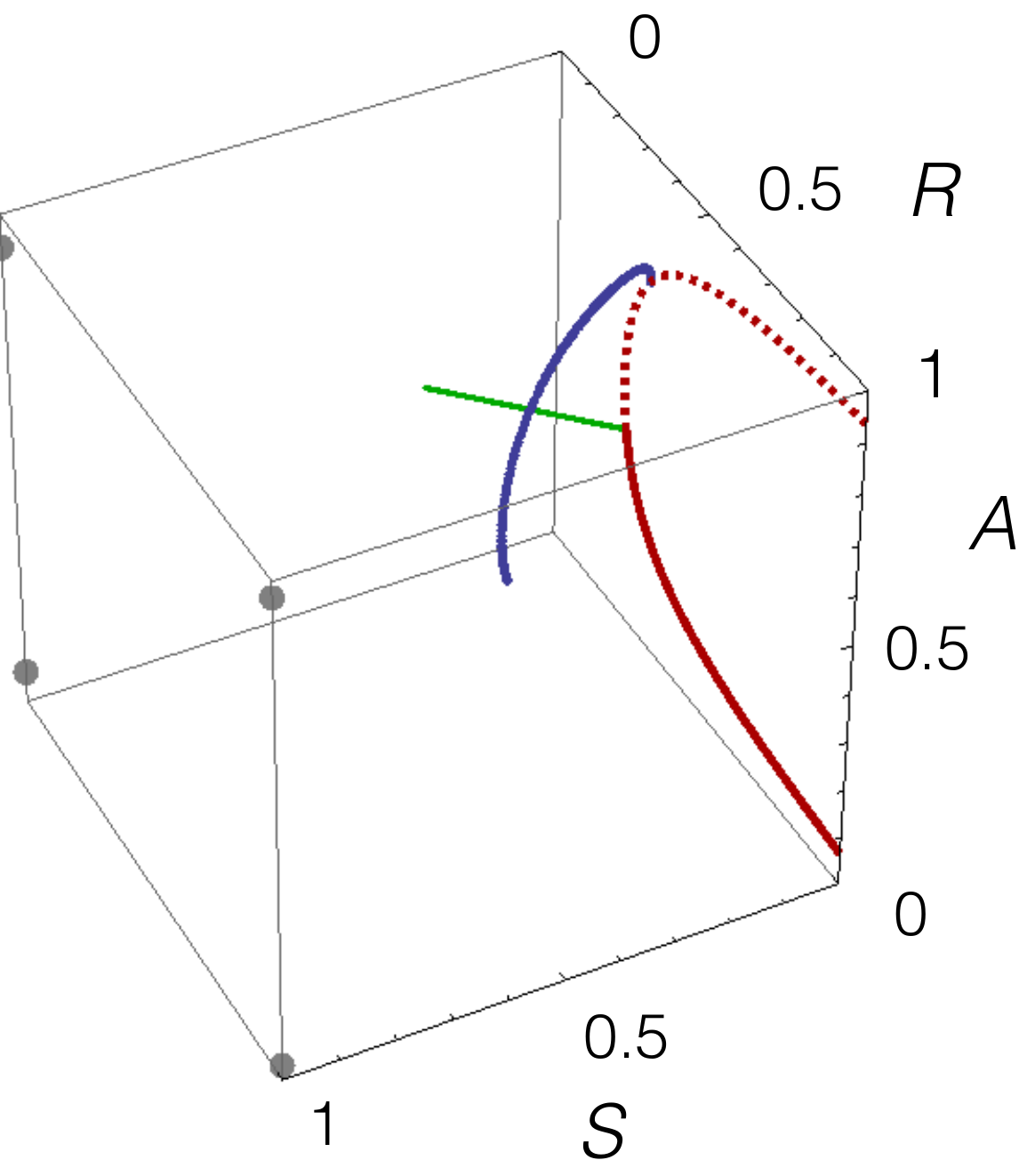}
\caption{The dynamics of the model for the case of the line (\ref{stableline}) being the repulsor.  The blue line is a trajectory of polity evolution of the system with parameter values:
$h=1$, $c=1$, $f=1$, $b=d=0.2$, $r=q=0.25$, and  $a=0.03$.  The initial conditions are $S_0=0.3$, $R_0=0.1$ and $A_0=0.3$  for the figure on the left,  and  $S_0=0.3$, $R_0=0.4$ and $A_0=0.3$ for the right. }
\label{f5}
\end{figure}
%

\subsection{The case $2b < \frac{h}{2} < 2d$}

For this choice of the parameter values, the trivial fixed points (\ref{fxtrv})-(\ref{fxs0}) on the surface of the $(S,R A))$-cube are given identically as before.
By examining the phase space strajectory of the system starting very slightly off ($R \gtrsim 0$) from the repelling fixed point, $(S, R, A) = (S^\star-f \epsilon, -\epsilon, 2 b )$ with very small $\varepsilon$, we find it forms a smooth curve that reaches the $S=0$ surface hitting the curve (\ref{fxr0}).   When $2b$ and $2d$ are both sufficiently close to $h/2$, the variation of $A(t)$ on the curve is very small, and the trajectory is described approximately by the condition
$\frac{d }{d t} A=0$, $\frac{d}{d t}(S+f R)=0$, namely, 
\begin{eqnarray}
\label{turchinfx}
\label{repl2}
&&
\frac{c}{1+f}(S+f R) A \left( 1-\frac{1}{h} A\right) - a = 0 \, ,
\nonumber \\
&&
 r \left( 1- \frac{1}{2b} A\right)  (1-S) + f q  \left( 1- \frac{1}{2d} A\right) R (1-R) = 0 \, ,
\end{eqnarray}
%
%
%
The line $\{ A=2b, S = S^\star - f R \}$ already gives a good starting point to approximate the solution of (\ref{repl2}), indicating that this object is a close relative to the line (\ref{stableline}) that exists for the case of $b=d$.
Assuming the solution in the form $\{ 2b+\Delta A, S^\star+\Delta S, R \}$, and taking the leading order of $\Delta A$ and $\Delta S$, we obtain a quite reasonable approximation
\begin{eqnarray}
\label{repap}
&&
A_r(R) = 2b +2b\frac{d-b}{d} Q(R) \frac{1}{1+\frac{b}{d}Q(R)}, 
\nonumber \\
&&
S_r(R) = -f R+S^\star - S^\star \frac{h-4b}{h-2b} \frac{d-b}{d} Q(R) \frac{1}{1+\frac{b}{d}Q(R)}
\label{repx}
\end{eqnarray}
with
\begin{eqnarray}
Q(R)= \frac{f q}{r} \frac{R(1-R)}{1-S^\star+f R} ,
\end{eqnarray}
for the axis-like object,
which is an invariant surface of codimension two.  Numerical examination of trajectories near the axis immediately reveals that they all move away from the axis.  Namely, we find this axis to be a repeller of the system, although it is not a collection of fixed points like in the case $b=d$.  
%

%
\begin{figure}[h]
  \centering
  \includegraphics[width=4.2cm]{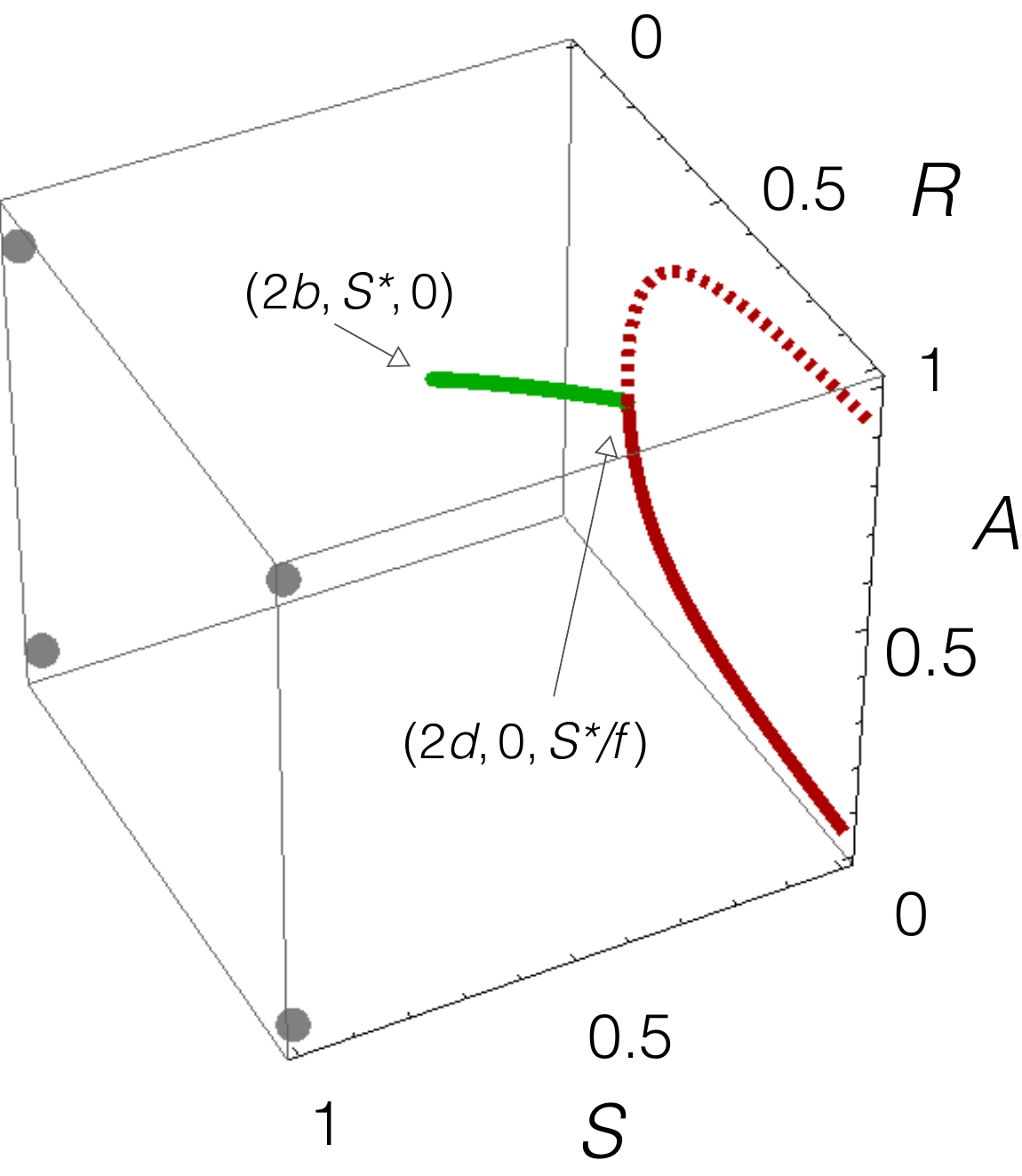}
\caption{
Invariant surfaces of the system with $2b < \frac{h}{2} < 2d$.  The grey points are the hyperbolic fixed points.  The solid and dotted red lines represent dense collections of repulsive and attractive fixed points, respectively.  The green line is the invariant surface from which orbits are repulsed.}
\label{f6}
\end{figure}
%
%
%
\begin{figure}[h]
  \centering
  \includegraphics[width=4.4cm]{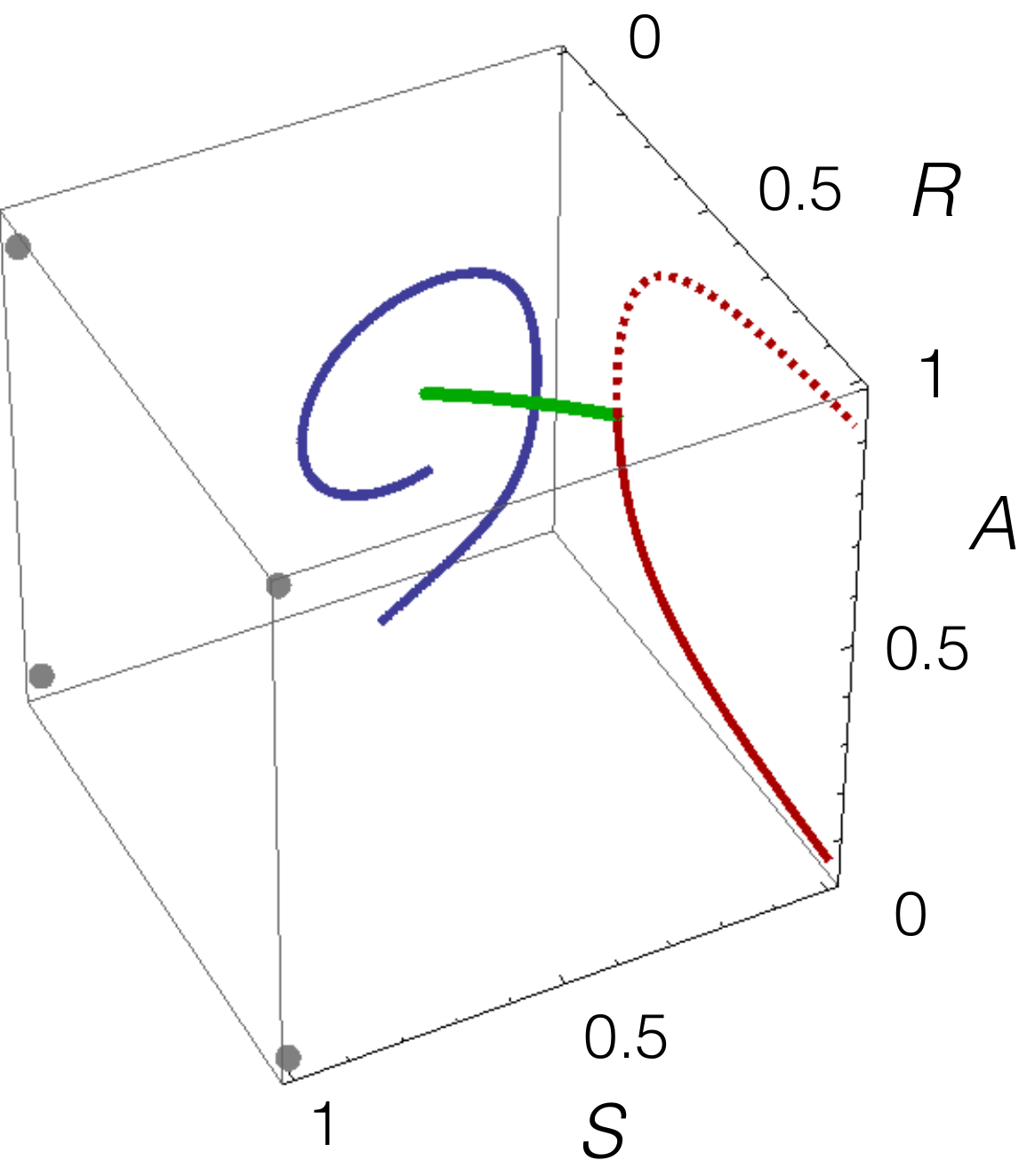}\qquad\qquad
  \includegraphics[width=4.4cm]{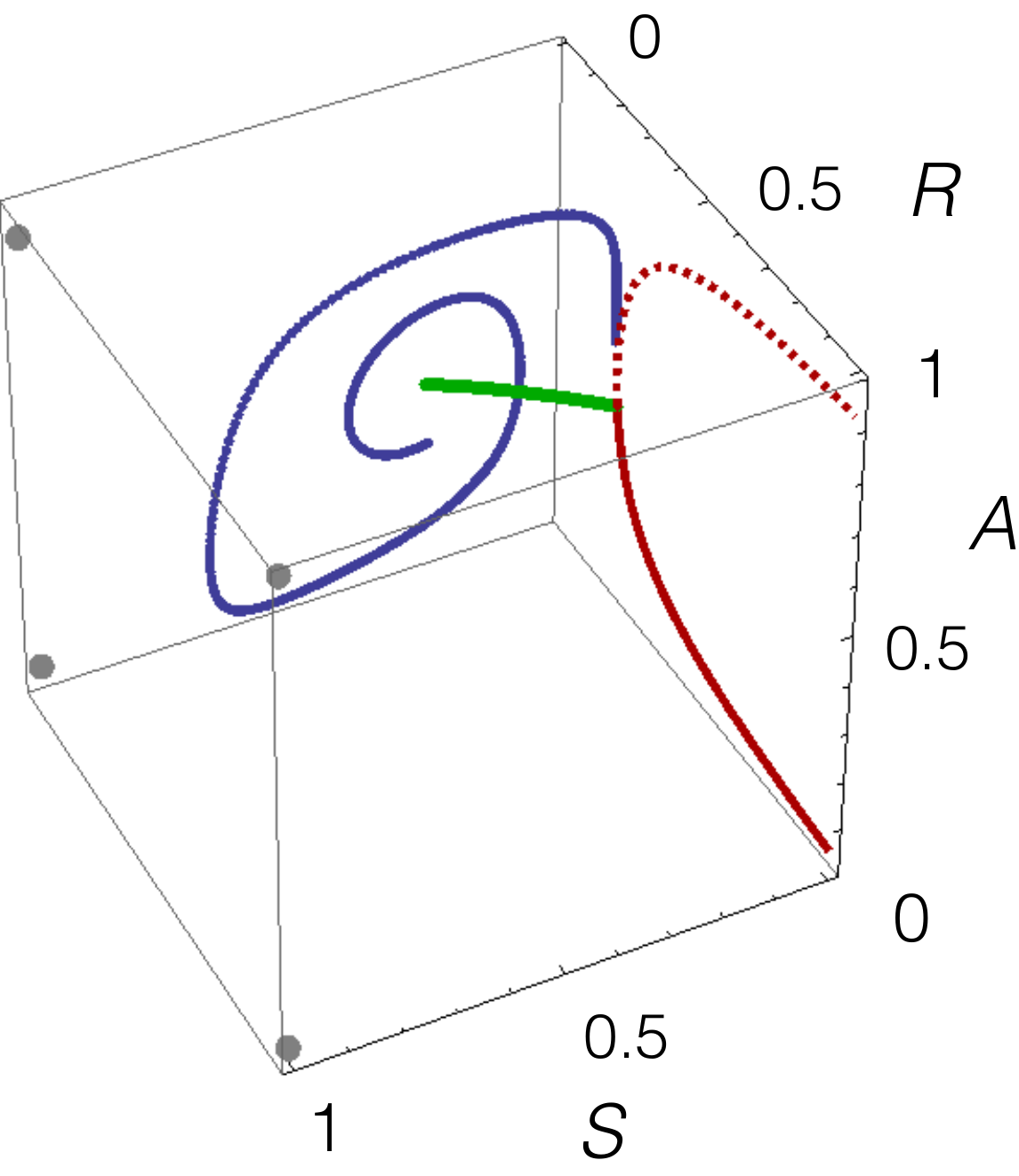}\\
  \bigskip
  \includegraphics[width=4.4cm]{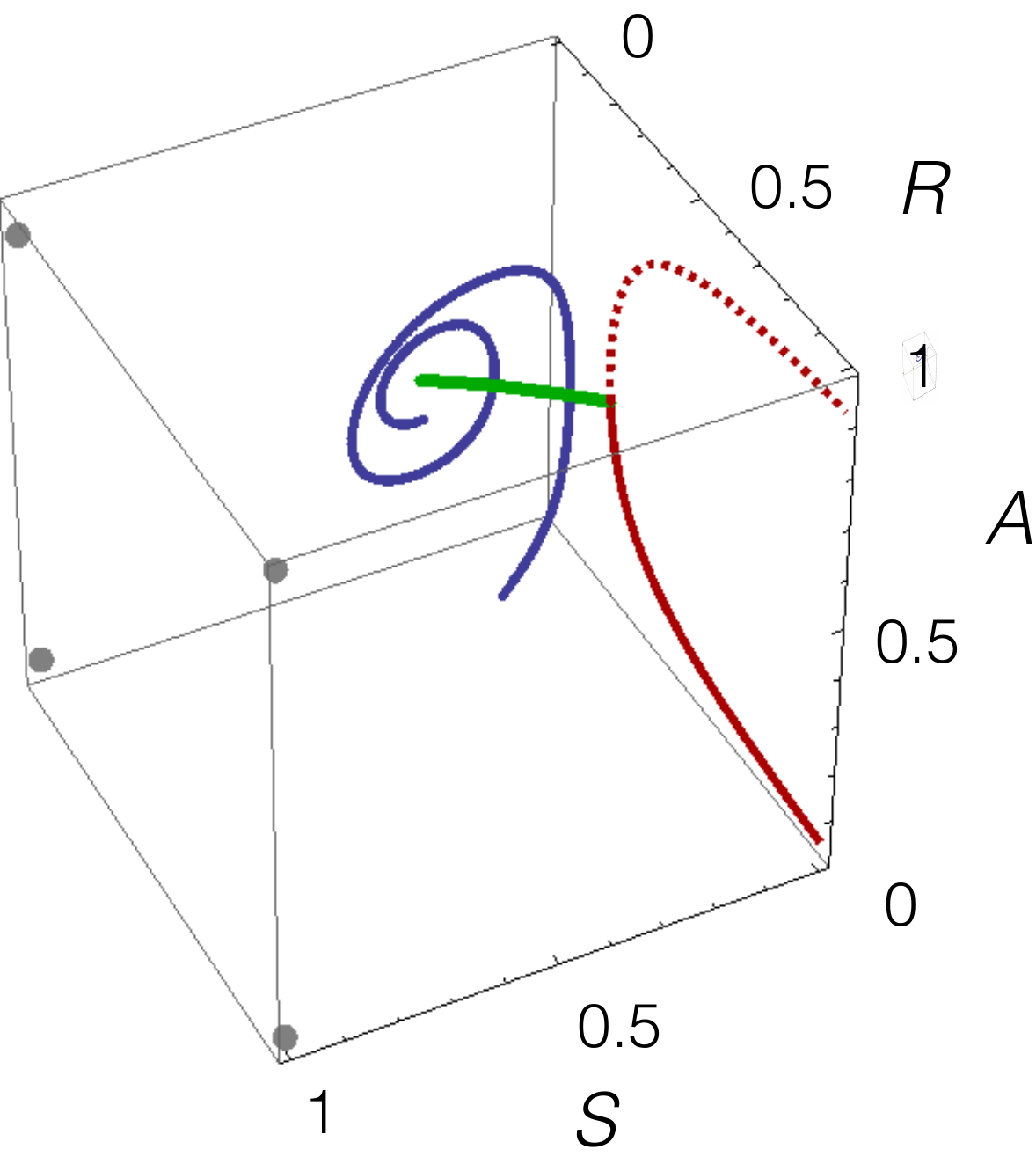}\qquad\qquad
  \includegraphics[width=4.4cm]{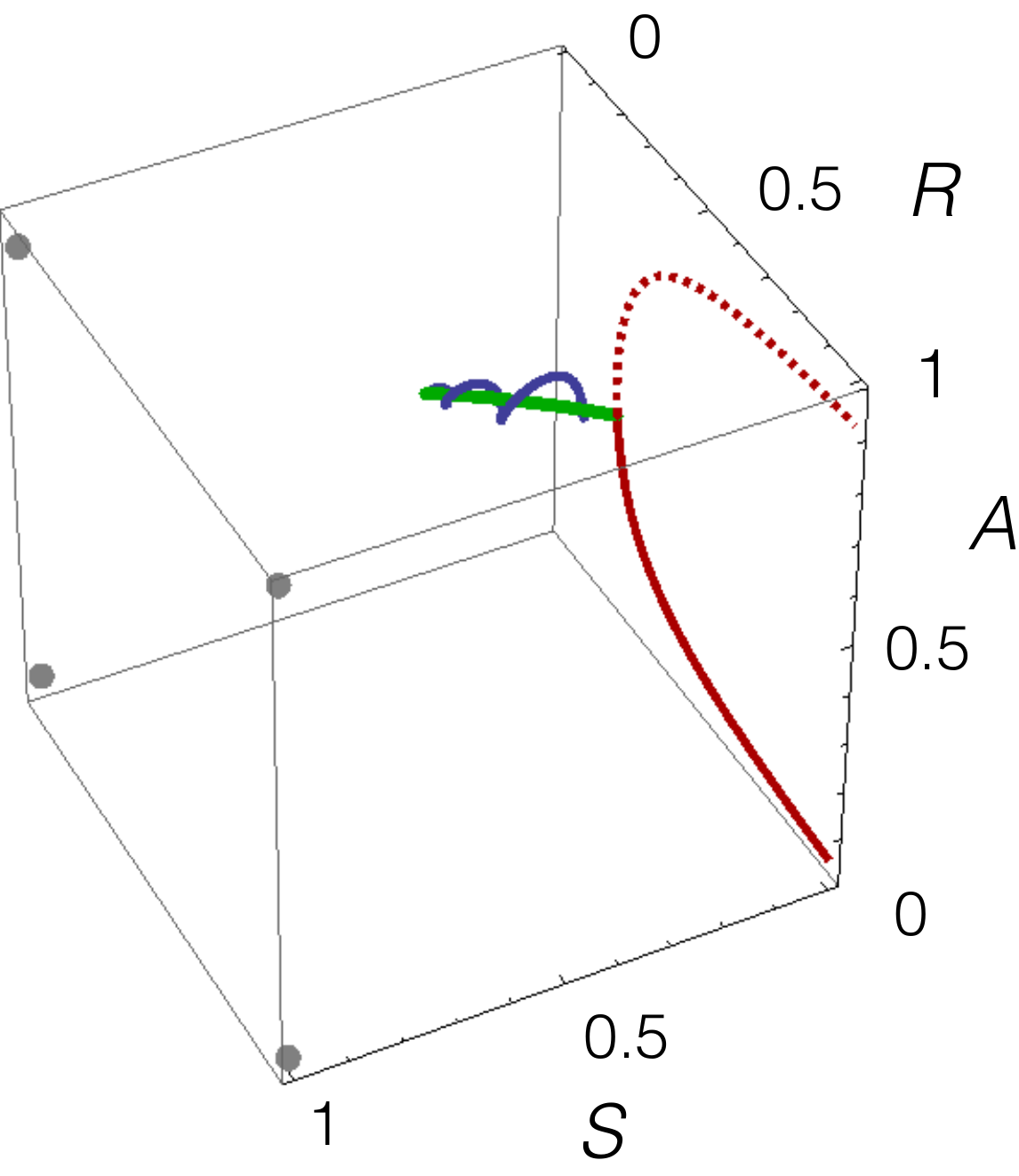}
\caption{ 
The evolution of the system shown in figure 1 is recast on the phase space. The phase space orbits (blue lines) of the dynamical system described by (\ref{exturchin}) starting from the initials values $A(0) = 0.245$ (top left), $A(0) = 0.285$ (top right), $A(0) = 0.325$ (bottom left), and $A(0) = 0.405$ (bottom right). Green and red lines are the invariant surfaces of the dynamics, which are identical to Fig. 6.  The choice of parameters is the same as in Fig. 1, and  each orbit corresponds to the temporal evolutions of $A$, $S$, and $R$ depicted in corresponding positions in Fig. 1. 
}
\label{f7}
\end{figure}
The skeletal elements of the dynamical system, namely, the elliptic fixed points, the attractor string and the repeller axis, are shown in Fig. 6 for the case of the specific parameter choice as before, $c=1$, $f=1$, $h=1$, $a=\frac{3}{100}$, $r=q=\frac{1}{4}$, $b=\frac{1}{5}$ and $d = \frac{2}{5}$.

The existence of repeller axis and attractor string is a distinctive feature of our system that comes from the particular combination of  a negative constant term and the third order terms in the model equations (\ref{exturchin}). 
The manner of spiral rotation around this axis characterizes the system's temporal evolution.
We depict, in Fig. 7, four orbits that correspond to the four examples of system evolutions given in Fig. 1, 
each of which we can identify as the single rise and fall, the fall after two oscillations, the ``eternal'' persistence after oscillations, and the approach to the persistence with mild oscillations.
These examples clearly show that all orbits rotate around this axis while moving from $R=0$ point to $S=0$ point, thus forming spirals around the repeller axis. 
We can observe that the repeller axis is working as the ``driving force'' of the rise and fall of polities, and also that the attractor string, whose all points are attractors, make the final settling points of polities unpredictable.  

For the case of $2b,2d<h/2$, and  $2b,2d>h/2$  with $b \ne d$, the situation is similar to the case of $b=d$, apart from the fact that the axis, either attracting or repelling, is specified by (\ref{repx}). 

\section{Analysis of spiral rotation around the repeller}

With the observation of the structure of the orbits around the repeller in numerical examples, such as in the one shown in Fig. 8, we can discern the phase space trajectory $\{ A(t), S(t), R(t) \}$ as a combination of the spiral confined mostly in $A-S$ plane and its procession along the repeller axis.
%
\begin{figure}[h]
  \centering
  \includegraphics[width=4.2cm]{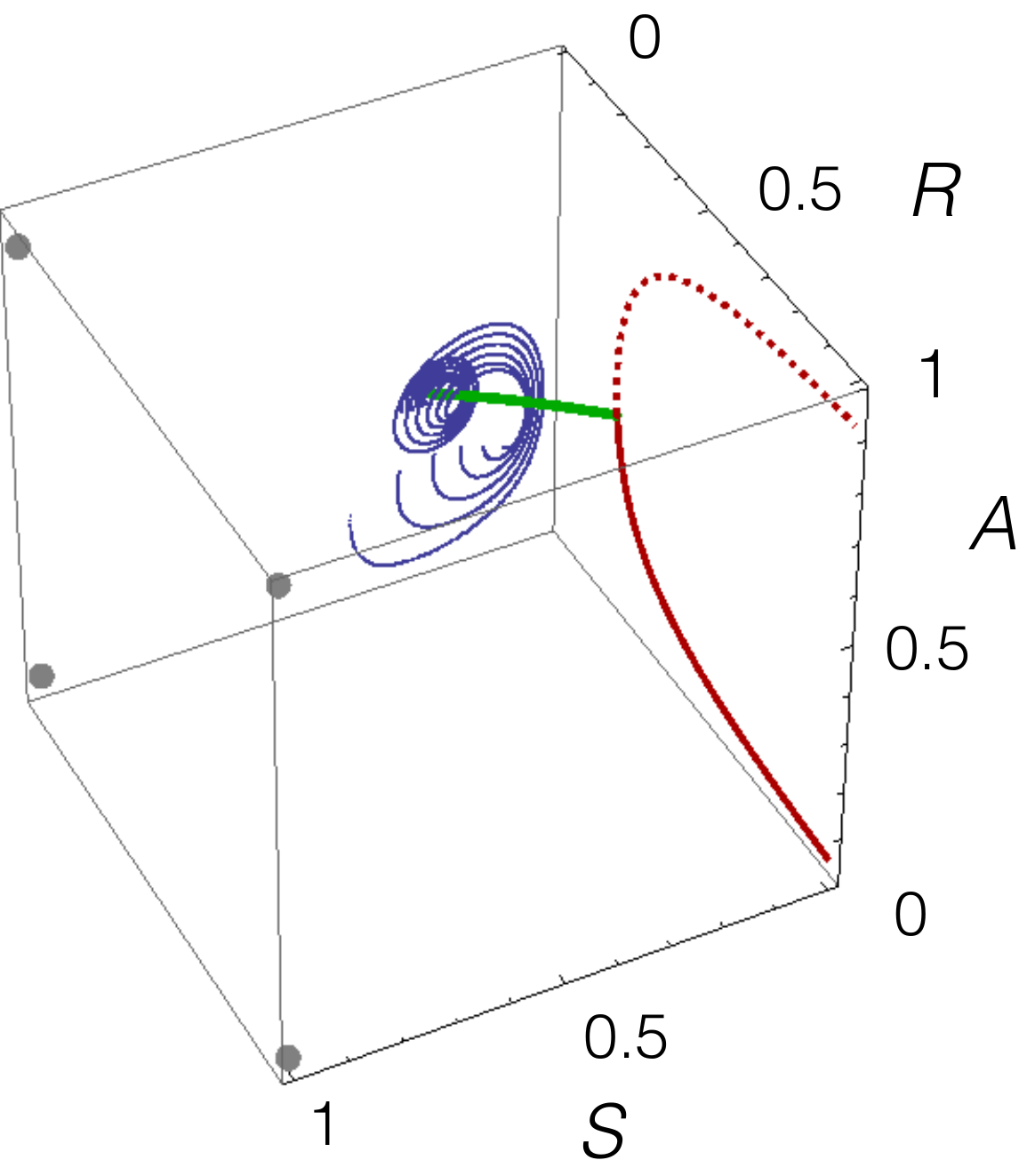}\qquad\qquad
  \includegraphics[width=4.2cm]{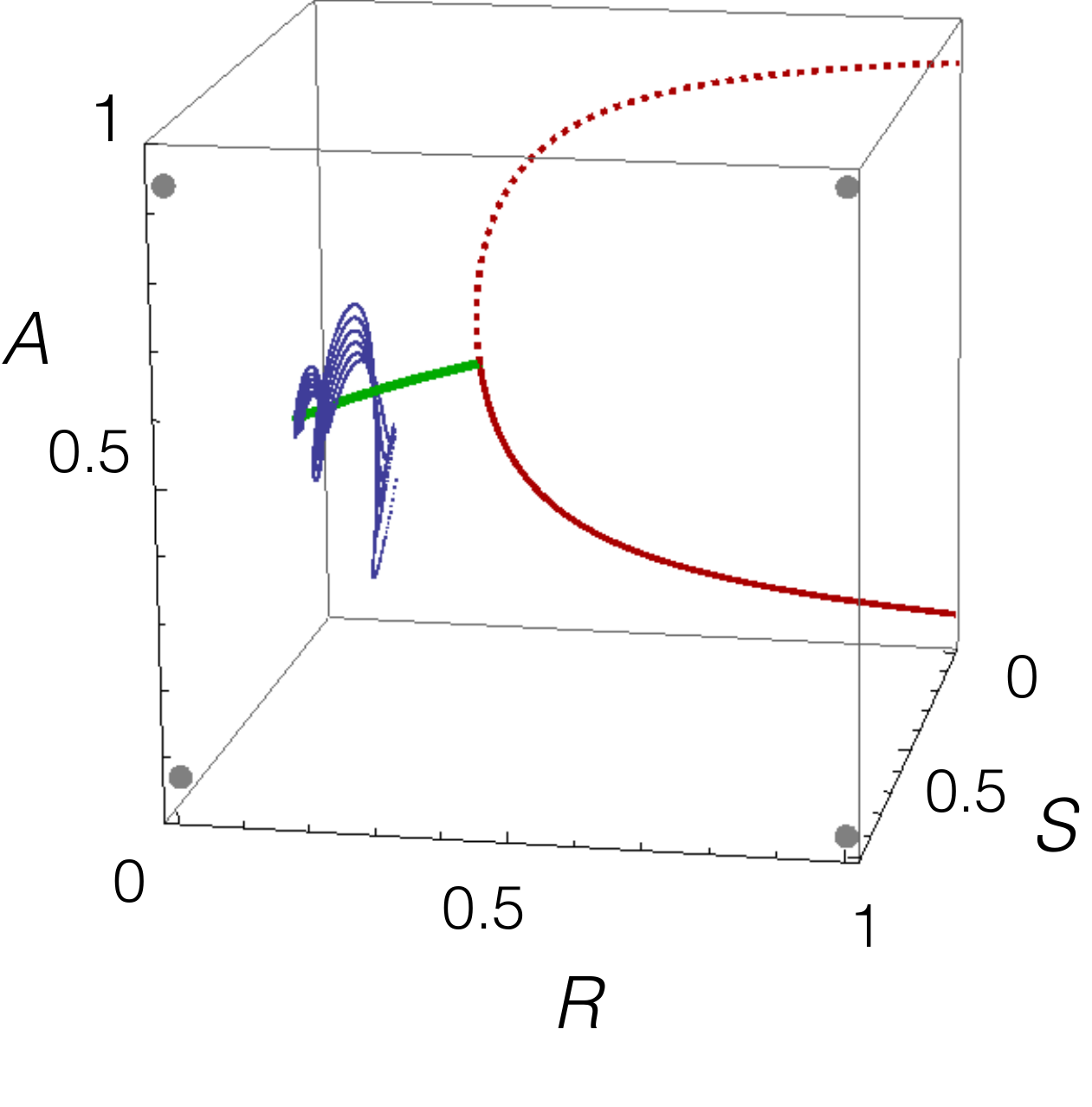}
\caption{ 
Six phase space orbits of the dynamical system described by (\ref{exturchin}) starting from the initials values $A(0) = 0.385$ and $S(0)$ ranging from  $0.25$ to $0.30$.  The choice of parameters are as in Fig. 1.  The left figure (a) and the right figure (b) are drawn from two different viewing angles. The trajectories for the time period $T=121$ are shown.
}
\label{f8}
\end{figure}
As a rough estimate for the spiral motion, let us consider the orbits near the $R=0$ edge of the repeller (\ref{repl2}), namely the fixed point $\{ 2b, S^\star, 0\}$ where $S^\star$ is given by (\ref{fxs0}).
Assuming that the orbit starts at $A$ and $S$ close enough to the fixed points, we expand the variables $A$ and $S$ in the form
$A=2b+\delta A$, $S=S^\star+\delta S$ 
and obtain, with the omission of $o(\delta A)$ and $o(\delta S)$ terms, the linearized map
\begin{eqnarray}
\frac{d}{dt}  \begin{pmatrix} \delta A \\ \delta S \end{pmatrix}
= \begin{pmatrix} \alpha & \beta \\ -\gamma & 0 \end{pmatrix}
\begin{pmatrix} \delta A \\ \delta S \end{pmatrix}
\end{eqnarray}
with
\begin{eqnarray}
\alpha = \frac{c (h-4b)}{(1+f) h} S^\star ,
\quad
\beta = \frac{a}{S^\star} ,
\quad
\gamma = \frac{ar}{2b} S^\star (1-S^\star) ,
\end{eqnarray}
Two quantities $\alpha$ and $\beta$ are always positive, and the third one $\gamma$ is also positive for sufficiently small $\lambda$. 
We easily obtain the equation for both $\delta A$ and $\delta S$ in the form
\begin{eqnarray}
&&
\frac{d^2}{dt^2}  (\delta S) - \alpha \frac{d}{dt}  (\delta S) + \omega_0^2 (\delta S) = 0 ,
\nonumber \\
&&
\frac{dA}{dt} = -\gamma S \, ,
\end{eqnarray}
with
\begin{eqnarray}
\omega_0^2 = \beta \gamma = \frac{ar}{2b}(1-S^\star), 
\end{eqnarray}
showing that the motions of $S(t)$ around $S^\star$ and $A(t)$  around $2b$ are both the expanding oscillations $\delta S(t) = \zeta_0\, {\rm e}^{\alpha/2} \cos (\omega t +\eta_0)$ and 
$\delta A(t) = -\gamma \zeta_0\, {\rm e}^{\alpha/2} \cos (\omega t +\eta_0 + \eta)$ with constants $\zeta_0$, $\eta_0$ specifying the initial conditions, and the frequency $\omega$ and the phase difference $\eta$ given by
\begin{eqnarray}
\label{e10}
&&
\omega=\sqrt{\omega_0^2-\alpha^2/4} \approx \omega_0 -\frac{\alpha^2}{8\omega_0} ,
\nonumber \\
&&
\eta = \cot^{-1} \frac{\alpha}{2 \omega_0} \approx \frac{\pi}{2} - \frac{\alpha}{2\omega_0} .
\end{eqnarray}
For the case of $\frac{\alpha}{2\omega} \ll1$ (for the parameter sets of our numerical example of Figs. 1 and 7, this is a quantity around 0.05), we have the phase difference of roughly $\frac{\pi}{2}$ between the $\delta S$ and $\delta A$ oscillations, making the $(A(t), S(t))$ motion expanding spirals around the repeller axis.
The period of oscillation $T_{osc} \approx \frac{2\pi}{\omega_0}$, which takes the form
\begin{eqnarray}
T_{osc} = \sqrt{ \frac{8 \pi^2 b} {ar (1-S^\star) } } ,
\end{eqnarray}
gives an estimate for a duration of time in which the polity experiences a single oscillating rise and fall in its population.  

As we can easily grasp from (\ref{exturchin}), the rate of change of metaasabiya $R$ is always positive, i.e. $R<2d$, for all orbits near enough to the repeller axis $0<R<S^\star/f$.  In fact, from Fig. 1 we observe that the metaasabiya increases steadily, when it starts from near zero value. However, it experiences a wild surge during the recovery period of the polity's ``second rise''.
We can make an estimate for the average rate of the increase of $R$ by calculating the average $\frac{dR}{dt}$ along the orbit on the repeller axis in the form
\begin{eqnarray}
\left< \frac{dR}{dt}\right> = \frac{1}{R_1}
\int_0^{R_1} dR q \left(1-\frac{A_r(R)}{2d}\right) S_r(R) R(1-R) ,
\end{eqnarray}
where $R_1 = \frac{S^\star}{f}$ is the saturation value of metaasabiya when the polity reaches the permanence.  Taking only the leading terms for $A_r(R)$ and $S_r(R)$ in (\ref{repap}), we obtain an excellent approximation $\frac{q (d-b))}{6fd} (S^\star)^2$ for this quantity.  The time necessary to reach the saturation value $T_R$ is estimated using the relation $T_R \left< \frac{dR}{dt}\right> = R_1$, for which we obtain
\begin{eqnarray}
T_R = \frac{ 6d }{ q(d-b) S^\star } .
\end{eqnarray}

Although these estimates are obtained with a restrictive assumption of the orbit being close enough to the repeller axis, the numerical calculation indicates that it gives a very reasonable guess for more general cases.  
For example, for the parameter given in the example of Fig. 1, we obtain the number $T_{osc}=53.0$ and $T_R=192$, which are quite satisfactory, in comparison to the value $T_{osc}\approx 60$ and $T_R \gtrsim 160$ estimated from the browsing of graphs in Fig. 1.
The quantities $T_{osc}$ and $T_{R}$ are determined solely by system parameters,  independently of initial conditions of orbits.   It can thus be utilized in the empirical estimations of parameter values from the historical data.

\section{Basins of attractions}
%
\begin{figure}[h]
  \centering
  \includegraphics[width=4.4cm]{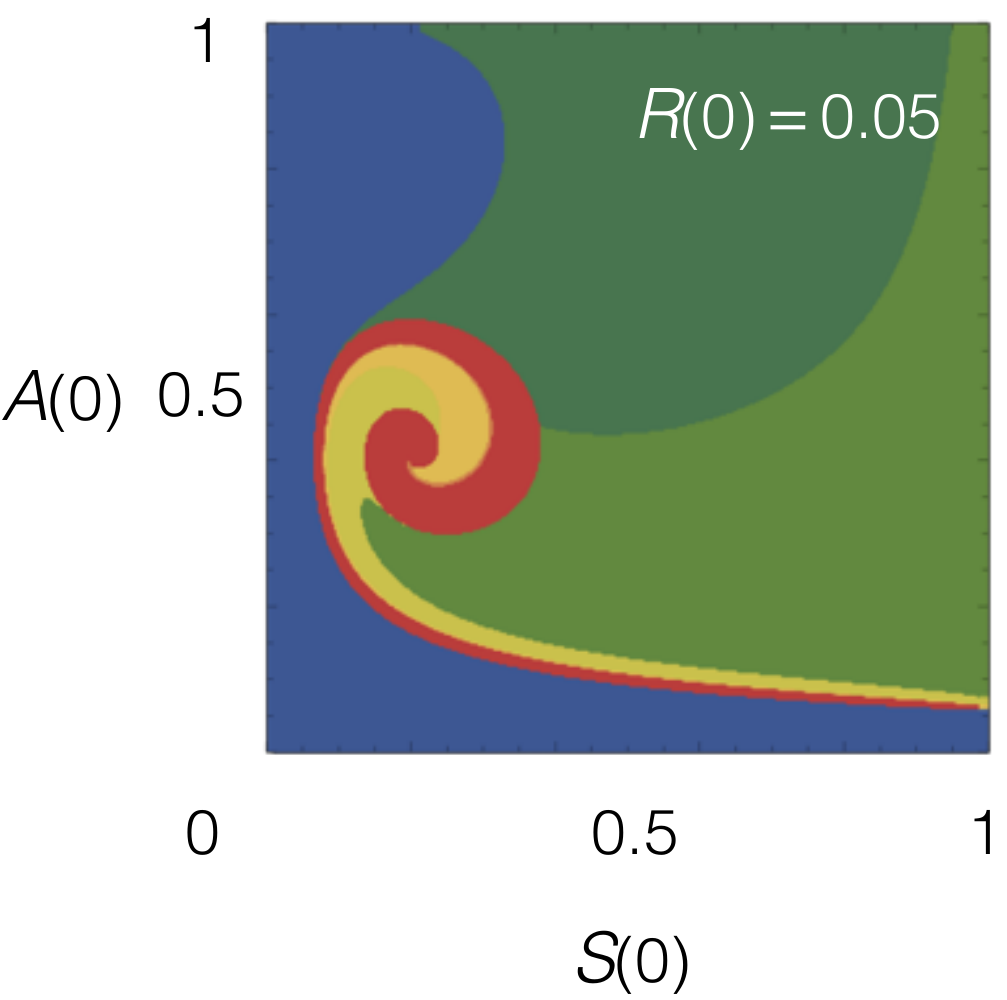}\qquad
  \includegraphics[width=4.4cm]{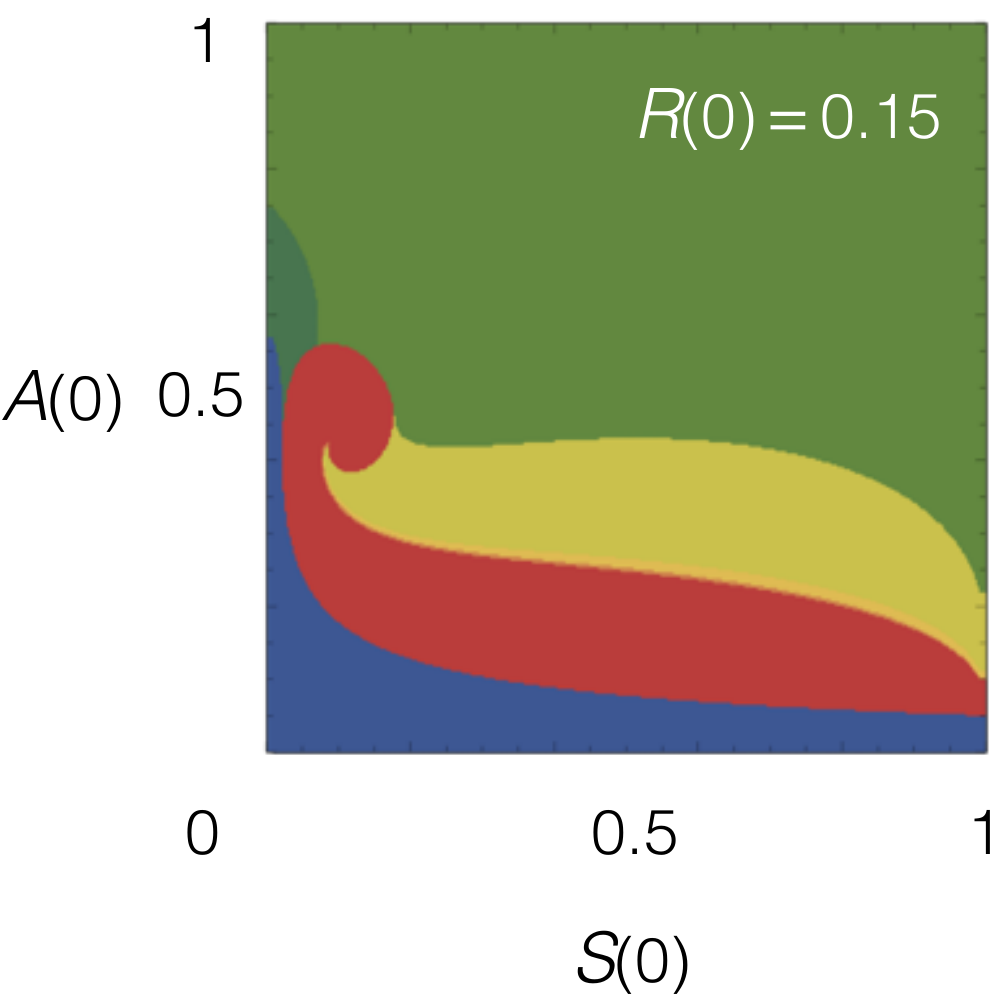}
\caption{
The basins of attraction in $\{ A(0),S(0)\}$ plane for the $A(t)\ne 0$ final state (red), for the eventual fall $A(t)=0$ after two rotations around the repeller (yellow), and for the fall $A(t)=0$ after a single rise (green and yellow-green) for the initial metaasabiya values of $R(0)=0.05$ (left), and for $R(0)=0.15$ (right).  The parameters are chosen to be identical to the ones for Fig. 1:
$c=1$, $f=1$, $h=1$, $a=\frac{3}{100}$, $r=q=\frac{1}{4}$, $b=\frac{1}{5}$ and $d = \frac{2}{5}$.
}
\label{f9}
\end{figure}
%
%
%
Now that we have a dynamical system describing the time evolution of a polity, the natural question to ask is which initial configurations of a polity leads to its rise and fall, fall after oscillations, and its permanence.  Obviously, initial conditions located on, or very close to the repeller axis should reach the attractor at $R=0$ plane, thus attain the permanence.  But there are initial configurations far from the repeller axis which, after several rotations, happen to hit the attractor, instead of hitting the ground plane $A=0$.  To decide the fates of polity for various initial conditions, we need to resort to numerical calculation. 

In Fig. 9, showing the $S(0)-A(0)$ plane at $R(0)=0.05$ (left) and at $R(0)=0.15$ (right), we depict the basin of attraction for the permanence (red), for  the fall after two oscillations (yellow), and for the single rise and fall (green), outright fall without any rise (blue).
These ``Ying-Yang''-shaped maps show that the longevity of a polity depends on the very intricate combination of the initial values of polity size,  asabiya and metaasabiya.

\section{Discussions}

In this article, we have developed a mathematical formulation of Toynbee's theory of history in the form of extended Turchin's equation.
We have shown that, with the introduction of second asabiya, the cliodynamical systems acquire rich dynamical structures not found in Lotka-Volterra systems, for example, which result in an intricate pattern to the temporal evolution of polities, revealing the hidden power inherent in the asabiya models.   The dynamical 
A distinct advantage of our three variable system is its capability of describing {\it varied fates} of a polity with different initial states under  {\it a single set of system parameters}.  In comparison, the original two-variable system requires a different number for the key parameter $b$ for the different final states. 
In hindsight, the result is very natural because the dynamics of two variable system is too limited to bring the diversity necessary to describe varied historical contours of polities.  As is well known, a dynamical system with three variables is capable of possessing such features as limit cycles and strange attractors \cite{AS96}.   It should be of much interest to search for the model with those properties.

Relating asabiya and mesaasabiya to measurable quantities in polities should be an urgent task, if we are to make good of the claim of cliodynamics that the theory is scientific in the sense it is experimentally provable  and refutable.  The concepts of perpendicular and horizontal propagation of cultural heritage in mathematical theory of cultural evolution might give some clues.
Also, the attempts to derive asabiya dynamics studied here from a microscopic game-theoretic processes of interactions between individuals in a polity is a hugely challenging and possibly very rewarding project \cite{ME11}.

Temporal variation of asabiya and metaasabiya should imply the change of internal structure of a polity, which is obviously beyond the scope of the current model that represent the whole polity with a single quantity, the size $A$. 
Further extension of current model with the consideration of class structure by dividing the population into that of elites and commoners should be of much interest.  
Even with the ``realistic'' extensions, we believe it certain, that the key for the dynamics to have multiple oscillatory rise and fall depends on the presence of repeller axis structure as in this model.
One element still missing in our treatment is the effect of the contacts with other polities.  Further extending the model with plural polity with geographical contact, as in the works of Artzrouni and Komlos \cite{AK96}, and also of Turchin \cite{TU03}, should be fruitful.

The asabiya, the concept of non-material ``spiritual'' resource specific to the human assembly, has been the key to produce ``the axis of history'' in its dynamical phase space.  We might then pose a question whether the distinctive dynamics described by the system of equations (\ref{exturchin}) is really specific to humanity, or can be also found in other biological ecosystems, possibly in the assembly of primates, if not in the colony of bacteria.   This line of inquiry might open up a new vista for the study of the evolutionary biological  complexities.

\bigskip\bigskip
\noindent{\bf Acknowledgements}

This research was supported by the Japan Ministry of Education, Culture, Sports, Science and Technology under the Grant number 15K05216.
We thank Dr. Ondrej Turek for stimulating discussions.

\bigskip


\end{document}